\documentclass[prb,preprint]{revtex4}
\usepackage[dvips]{graphicx}
\usepackage{here}
\usepackage{natbib}
\newcommand{\beq}{\begin{equation}}
\newcommand{\eeq}{\end{equation}}
\newcommand{\mbold}[1]{\mbox{\boldmath $ #1 $}}
\begin{document}

\title{Critical temperature and correlation length of an elastic interaction model 
for spin-crossover materials}
\author{Taro Nakada$^{1}$, Takashi Mori$^{1}$, Seiji Miyashita$^{1,2}$, 
Masamichi Nishino$^{3}$, 
Synge Todo$^{2,4}$, William Nicolazzi$^{4,5}$, and Per Arne Rikvold$^{6}$}
\affiliation{
$^{1}${\it Department of Physics, Graduate School of Science,
The University of Tokyo, 7-3-1 Hongo, Bunkyo-Ku, Tokyo 113-8656, Japan.}  \\
$^{2}${\it CREST, JST, 4-1-8 Honcho Kawaguchi, Saitama 332-0012, Japan.}  \\
$^{3}${\it National Institute for Materials Science, Tsukuba, Ibaraki 305-0047, Japan.}  \\
$^{4}${\it Department of Applied Physics, University of Tokyo, Tokyo 113-8656, Japan.}
$^{5}${\it CNRS,Laboratoire de Chimie de Coordination (LCC/CNRS/UPR8241)
205, route de Narbonne, 31077 Toulouse cedex4, France.} \\
$^{6}${\it Department of Physics, Florida State University, Tallahassee, Florida 32306-4350, USA.}}

\begin{abstract}

It has previously been pointed out that the coexistence of infinite-range and short-range 
interactions causes a system to have a phase transition of the mean-field universality class, 
in which the cluster size is finite even at the critical point. 
In the present paper, we study this property in a model of bistable molecules, whose size changes 
depending on the bistable states. 
The molecules can move in space, interacting via an elastic interaction. 
It is known that due to the different sizes, an effective long-range interaction between the 
spins appears, and thus this model has a mean-field type of phase transition. 
It is found that the scaling properties of the shift of the critical temperature from 
the pure short-range limit in the model with infinite-range and short-range 
interactions hold also in the present model, 
regarding the ratio of the size of the two states as a control parameter for the strength of the 
long-range interaction. 
By studying the structure factor, it is shown that the dependence of the cluster size 
at the critical temperature also shows the same scaling properties as a 
previously studied model with both infinite-range and short-range interactions. 
We therefore conclude that these scaling relations hold universally 
in hybrid models with both short-range and weak long-range interactions. 
\end{abstract}

\date{\today}
\maketitle

\section{Introduction}
The divergence of the correlation length at the critical point is 
considered to be one of the most important properties of 
second-order phase transitions. 
However, it is also known that 
in phase transitions of infinite-range interacting systems that belong to 
the mean-field universality class, the correlation length does not 
diverge, and the spatial configuration is uniform with 
no domain structures or clustering. 
In a previous work, we studied a hybrid model 
with both short-range and weak long-range interactions.\cite{Nakada} 
The Hamiltonian of that model is given by
\beq
{\cal H}=(1-\alpha){\cal H}_{\rm IS}+\alpha{\cal H}_{\rm MF}, 
\hspace{1cm}(0\le \alpha \le 1)
\label{model2}
\eeq
with
\beq
{\cal H}_{\rm IS}=-J\sum_{\langle i,j \rangle}\sigma_i \sigma_j,
\label{modelis}
\eeq
and
\beq
{\cal H}_{\rm MF}=-{4J\over 2N}\sum_{ i,j }\sigma_i \sigma_j=-{2J\over N}\sum_{ i,j }\sigma_i \sigma_j.
\label{modelmf}
\eeq
We defined the model on the square lattice with periodic boundary conditions. 
Here, the strength of the infinite-range 
interaction is controlled by varying $\alpha$. 
When $\alpha=0$, the system is equivalent to the pure 
short-range Ising model, and the system with $\alpha=1$ is equivalent to the 
pure infinite-range interaction model. 
The critical temperature in the previous hybrid model (\ref{model2}) shows a crossover from 
that of the pure short-range Ising model to 
that of the infinite-range interaction model.  
It should be noted that even for infinitesimally small $\alpha$, 
the phase transition belongs to the mean-field universality class, and at the 
critical point, the spin configuration is uniform with no 
large-scale clustering. 
A scaling formula for the $\alpha$-dependence of the critical 
temperature $T_{\rm c}$ is found, such that 
\beq
{T_{\rm c}(\alpha)-T_{\rm c}^{\rm IS}\over T_{\rm c}^{\rm MF}-T_{\rm c}^{\rm IS}}\simeq 
1.773517\alpha^{1\over \gamma}
=1.773517\alpha^{4\over 7}.
\label{tcprevious}
\eeq
The correlation length $\xi_{\rm c}$ at $T_{\rm c}$ 
is also found as a function of $\alpha$, 
\beq
\xi_{\rm c}(\alpha,L)=Lf(L\alpha^{\nu\over \gamma})= 
Lf(L\alpha^{4\over 7})
\label{xiprevious}
\eeq
for small $\alpha$ and large $L$. 
The function 
$f(x)$ is a scaling 
function which asymptotically approaches $1/x$ for large $x$. 
\par
Although the long-range interaction in this model is rather artificial, 
recently it has been pointed out that 
spin-crossover materials and 
related materials show a similar kind of 
long-range correlation.\cite{Miya1} 
Spin-crossover materials are molecular crystals, 
in which the molecules can exist in two different states: 
the high-spin (HS) state and the low-spin (LS) state.
The HS state is preferable at high temperatures 
because of its high degeneracy, while the LS state 
is preferable at low temperatures because of its 
low enthalpy. 
In addition to temperature, 
pressure changes and light exposure also 
often induce a phase transition 
in spin-crossover materials. 
Spin-crossover and related materials are used in many applications, 
because of their inherent bistability 
that leads to changes in optical and magnetic properties, etc. \cite{Ronayne}
Phase transitions in spin-crossover and related materials 
have been studied extensively in chemistry 
\cite{Gutlich,Real,Kahn,Letard,Hauser,Shimamoto,Molnar} 
and recently 
also in physics.\cite{Nishino,Enachescu,Nishino2,Konishi,McDonald,Huby} 
For a wide variety of applications, it is of great interest to study the ordering process 
in spin-crossover materials. 
\par
This type of materials are also regarded as fundamental models for 
inter-molecular short-range and 
elastic interactions due to the lattice distortion, 
and an elastic interaction model has been proposed. \cite{Miya1,Nishino} 
An important characteristic of this model 
is an effective long-range interaction 
due to the lattice distortion caused by the size difference between the HS (large) and LS (small) molecules. 
Even at the critical point, there exists no large-scale domain structures. 
However, in the elastic interaction model, 
there does not exist inter-molecular short-range interaction, 
so the model does not show any crossover from 
an effective short-range interacting system to an effective long-range 
interacting system.  
Therefore, developing quantitative hybrid models with both short-range and 
the elastic interactions for the critical behavior of such materials 
is very important. 
\par
In this paper, we focus on a system 
with both elastic interactions 
and short-range Ising interactions. 
We perform Monte Carlo (MC) simulations to study the properties of this model at the 
critical point. 
The rest of the paper is organized as follows. 
In Sec. \ref{secsec}, we introduce 
a new model with both elastic and 
short-range Ising interactions, 
and we propose a relation between this model and 
the model studied previously. \cite{Nakada} 
We find that the scaling formulae for the critical temperature and 
correlation length obtained in the previous paper \cite{Nakada} apply to this model, as well.
In Sec. \ref{secthir}, 
we briefly review the MC algorithms and 
give the result of MC simulations for the critical temperature, 
confirming the relation.
In Sec. \ref{secfour}, 
we similarly confirm the scaling form of the correlation length 
at the critical point. 
In Sec. \ref{secfif}, we summarize our results, and 
in appendix~\ref{app} we discuss in detail 
how we calculate the correlation length at the critical point.
\section{Model}
\label{secsec}
In this paper, we adopt the following model with both lattice distortion and 
inter-molecular short-range interactions on the square lattice 
with periodic boundary conditions. 
From now on, $L$ denotes the number 
of molecules along an edge of the lattice, 
so the total number of sites is $N=L^2$. 
The Hamiltonian is 
\beq
{\cal H}={\cal H}_{\rm IS}+{\cal H}_{\rm nn}+{\cal H}_{\rm nnn}+{\cal H}_{\rm eff},
\label{model3}
\eeq
with
\begin{eqnarray}
\left\{
\begin{array}{lll}
{\cal H}_{\rm IS}=-J\sum_{\langle i,j \rangle}\sigma_i\sigma_j\\
{\cal H}_{\rm nn}={k_1 \over 2}\sum _{\langle i,j \rangle}
\left[| {\mbold r}_i-{\mbold r}_j |-\left(R_i(\sigma_i)+R_j(\sigma_j)\right)\right]^2,\\
{\cal H}_{\rm nnn}={k_2 \over 2}\sum _{\langle\langle l,m \rangle\rangle}\left[| {\mbold r}_l-{\mbold r}_m |-\sqrt{2}\left(R_l(\sigma_l)+R_m(\sigma_m)\right)\right]^2,\\
{\cal H}_{\rm eff}=\left(D-{k_{\rm B}T\over 2}\log g\right)\sum_i^{N} \sigma_i,
\label{modeleff}\\
\end{array}
\right.
\end{eqnarray}
where ${\mbold r}_i$ represents the continuous coordinate of the molecule $i$, and $R_i(\sigma_i)$ is the radius of the molecule $i$ which depends on the molecular state $\sigma_i$. 
Here $\sigma_i=+1$ and $-1$ represents HS and LS state, respectively. Hereafter, we simply call $\sigma_i$ the spin state. 
${\cal H}_{\rm IS}$ is the short-range pure Ising model. 
${\cal H}_{\rm nn}$ and ${\cal H}_{\rm nnn}$ denote the elastic interaction 
Hamiltonians of nearest-neighbor $\langle i,j \rangle$ and 
next-nearest-neighbor $\langle\langle l,m \rangle\rangle$ 
pairs, respectively, and 
$k_1$ and $k_2$ are the corresponding spring constants. 
The next-nearest-neighbor interaction is introduced to maintain 
the shape of the lattice, and the strength of $k_2$ is not 
important as long as the global shape of the square lattice is kept. 
Here we take $k_2=k_1/10$. 
We define the pure elastic interaction model as 
\beq
{\cal H}_{\rm Elastic}\equiv{\cal H}_{\rm nn}+{\cal H}_{\rm nnn}.
\label{model1}
\eeq
The molecular radius is determined by the local spin state : $R_{\rm H}$ for the HS state 
(large, $\sigma_i=+1$) and $R_{\rm L}$ for the LS state (small,  $\sigma_i=-1$). 
When $| {\mbold r}_i -{\mbold r}_j |$ is equal to the sum of the radii $R_{i}(\sigma_i)+R_{j}(\sigma_j)$, 
the corresponding contribution to the elastic energy has its minimum. 
In ${\cal H}_{\rm eff}$, which represents the ligand field, $D$ denotes the energy difference between the HS state and LS state, 
and $g$ denotes the ratio of the degeneracies of the HS state and LS state. 
The order parameter of these models (\ref{model3}) and (\ref{model1}) is defined as
\beq
m\equiv{\sum_i^{N}\sigma_i\over N},
\eeq
which is related to the fraction of HS molecules, $f_{\rm HS}$, 
as $m=2f_{\rm HS}-1$. 
\par
In order to see the competition between the short-range interaction and the lattice distortion due to the molecular size difference, 
we consider the model (\ref{model3}) along the coexistent line, 
$\langle m \rangle=0$, as we studied in the previous work for the Ising model. 
Thus we set $D-(k_{\rm B}T/2)\log g=0$. 
For simplicity, this situation is described by the present model (\ref{model3}) in which $D=0$ and $g=1$. 
In this model, a ferromagnetic second-order phase transition 
takes place at the critical temperature $T_{\rm c}$. 
Below $T_{\rm c}$, there exist two different ordered states,
the HS state and the LS state. 
\par
It has been found that in the pure elastic interaction model~(\ref{model1}), 
the spin configuration is uniform at the critical point, with no 
large-scale clustering. \cite{Miya1} 
The phase transition belongs to the mean-field (MF) universality class, 
and the spin correlation function approaches a 
non-zero constant in the long-distance limit. 
\begin{figure}[H]
$$
\includegraphics[scale=0.5]{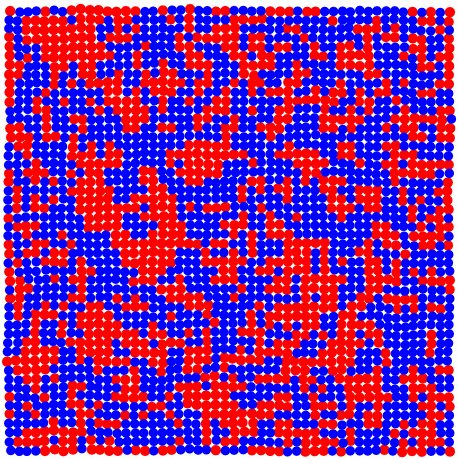}
$$
\caption{(Color online) Typical configuration of the mixed elastic and short-range interaction model at the critical point. 
$R_{\rm H}/R_{\rm L}=1.1$, for which 
$T_{\rm c}=0.54575$. 
Light gray (red online) filled circles denote HS state spins, and dark gray (blue online) filled circles denote 
LS state spins. 
}
\label{config}
\end{figure}

The origin of the long-range interaction in the 
elastic model is lattice distortion due to the 
size difference between the HS and LS states. 
We expect that 
the ratio of the radii $R_{\rm L}/R_{\rm H}$ 
controls the strength of the effective long-range interaction.
Namely, $R_{\rm L}/R_{\rm H}$ plays a role similar to $\alpha$ in 
the previous hybrid model (\ref{model2}). 
The strength of the long-range interaction is given 
by the elastic energy, which is of the order of 
$(k_1/2)(R_{\rm H}-R_{\rm L})^2\propto (1-{R_{\rm L}\over {R_{\rm H}}})^2$. 
We therefore consider that the parameter 
to indicate the strength of the long-range interaction is given by 
\beq
\alpha \propto \left(1-{R_{\rm L}\over R_{\rm H}}\right)^2.
\label{alpharadii}
\eeq
With this relation, we expect that the relations 
(\ref{tcprevious}) and 
(\ref{xiprevious}) take the following forms, 
\beq
T_{\rm c}\left({R_{\rm L}\over R_{\rm H}}\right)-T_{\rm c}^{\rm IS}\propto 
\left(1-{R_{\rm L}\over R_{\rm H}}\right)^{ 8\over 7},
\label{tc}
\eeq
and
\beq
\xi_{\rm c}\left({R_{\rm L}\over R_{\rm H}},L\right)= 
Lf\left(L\left(1-{R_{\rm L}\over R_{\rm H}}\right)^{8\over 7}\right),
\label{xi}
\eeq
respectively. 
As in our previous work,\cite{Nakada} 
as long as the value of 
$1-R_{\rm L}/R_{\rm H}$ is small, we expect 
these formulae to be correct. 
In Fig.~\ref{config}, we depict a typical configuration of 
the elastic and short-range interaction model 
at its critical temperature. 
Unlike the pure short-range Ising model, 
the spin configuration is uniform with no 
large-scale clustering, even at the critical temperature. 
A lattice distortion also occurs, and we observe an uneven system surface. 

\section{Critical temperature}
\label{secthir}
In this section, we perform MC simulations to 
test the scaling relation for the critical temperature~(\ref{tc}). 
For the simulation, we adopt the $NPT$-MC method \cite{McDonald} for 
the isothermal-isobaric ensemble with the number of molecules $N$, 
the pressure of the system $P$, and the temperature $T$. 
In this paper, in order to exclude other effects than those 
due to the elastic interaction through distortion, we fix $P=0$. 
We also fix the spring constants as $k_1=40$ and $k_2=4$ 
as in our previous work. \cite{Miya1} 
The critical temperature of the pure elastic interaction model (\ref{model1}) is 
$T_{\rm c}^{\rm Elastic}\simeq 0.2$ for $R_{\rm H}=1.1$, \cite{Miya1}
and we choose $J=0.1$ in order to keep the two terms in the present model (\ref{model3}) of comparable 
magnitude near the critical temperature. 
Therefore, $T_{\rm c}^{\rm IS}=0.2269\cdots$ in these units on the square lattice. 
We use a standard Metropolis method, adopting periodic 
boundary conditions. 
In most cases, we performed eight independent runs of 4,000,000 Monte Carlo steps per spin (MCSS) 
for the each data with 100,000 MCSS for the initial equilibration. 
We confirm that the statistical errors are smaller than the marks in the following graphs. 
We fix $R_{\rm L}=1.0$ and choose $R_{\rm H}=1.005, 1.008, 1.010, 1.015, 1.02, 1.05$, and $1.1.$ 
\\
We have previously pointed out that even infinitesimally weak 
long-range interactions become dominant in the thermodynamic limit. \cite{Nakada}
In the case of the previous hybrid model~(\ref{model2}), this property is explained by the fact that 
in a well coarse-grained Hamiltonian, the long-range 
interactions become stronger than the short-range ones. 
We need systems sufficiently large that 
clusters caused by the short-range interactions 
can be regarded as block spins. \cite{Nakada}
Here we assume that this size dependent crossover phenomenon also 
takes place in the present model~(\ref{model3}). 
\par
Here, in order to estimate the critical temperature, we adopt 
a method we also used in our previous paper. 
We use the crossing point of the forth-order 
Binder cumulant \cite{bindercumulant} $U_4(L)$ for different system sizes to estimate the critical temperature 
$T_{\rm c}(R_{\rm L}/R_{\rm H})$ with high precision. 
The cumulant is defined as
\beq
U_4(L)\equiv 1-{\langle m^4\rangle_L\over 3\langle m^2 \rangle^2_L},
\eeq
where $m=1/L^2\sum_i\sigma_i$. 
In the case of the pure elastic interaction model (\ref{model1}) 
on the square lattice with periodic boundary conditions, 
the fixed-point value of the cumulant \cite{Miya1} is the same as the exact value 
for the infinite-range interaction model (\ref{modelmf}), 
$U_{4}^{\rm MF}\simeq0.27\cdots$. \cite{Brezin,Luijten} 
With the radii ratio $R_{\rm L}/R_{\rm H}=1$, the present model (\ref{model3}) is equivalent 
to the pure Ising model on the square lattice with periodic boundary conditions. 
For this case the fixed-point value of the cumulant is the same as the value 
for the Ising model (\ref{modelis}), 
$U_{4}^{\rm IS}\simeq0.61\cdots$ on the square lattice. \cite{IsingCum} 
Other shapes of the system, boundary conditions, and anisotropy may lead 
to different values of $U_4$ at the crossing point. \cite{Chen, Selke}

\begin{figure}[H]
$$\begin{array}{cc}
\includegraphics[scale=0.4]{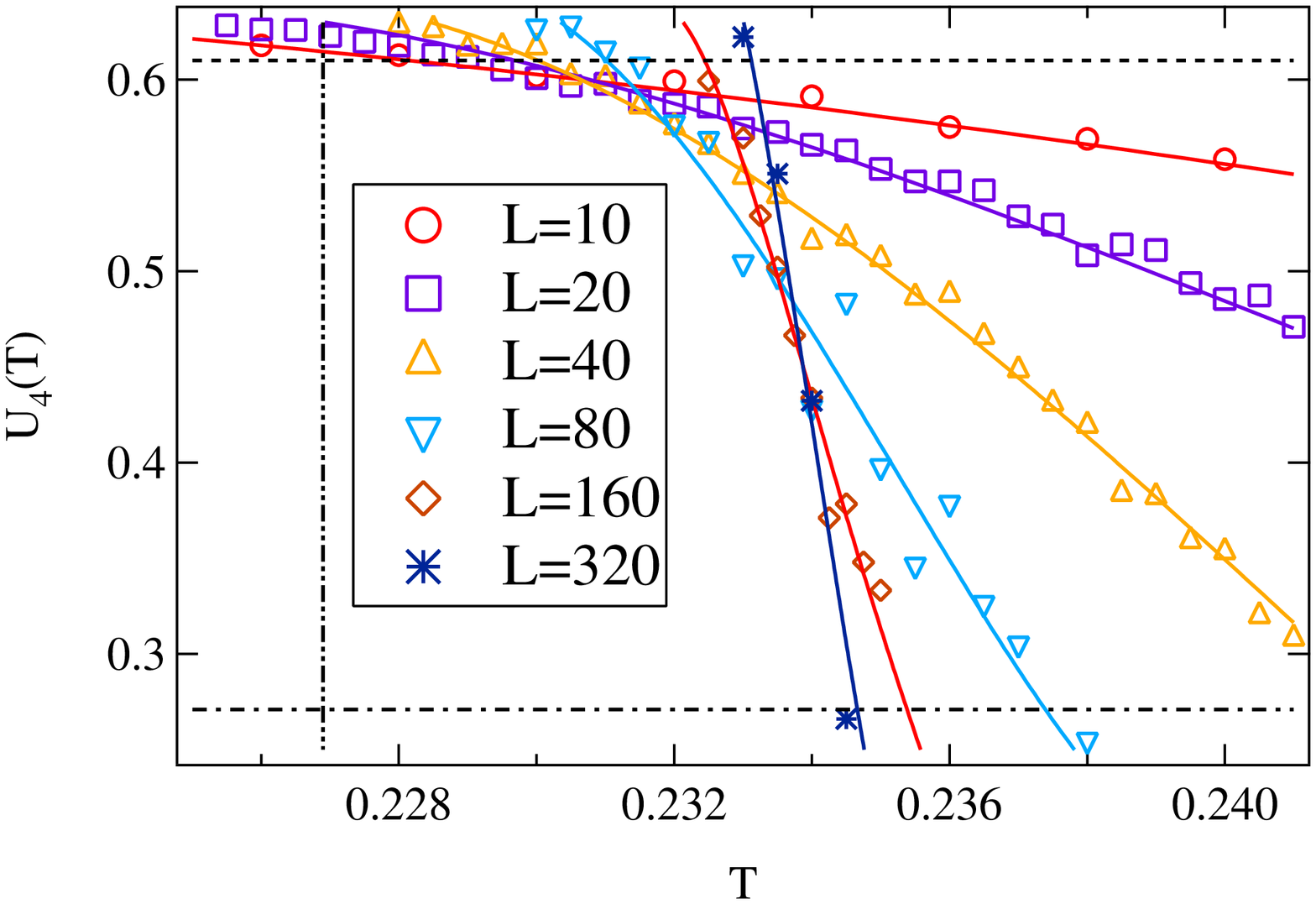} &
\includegraphics[scale=0.4]{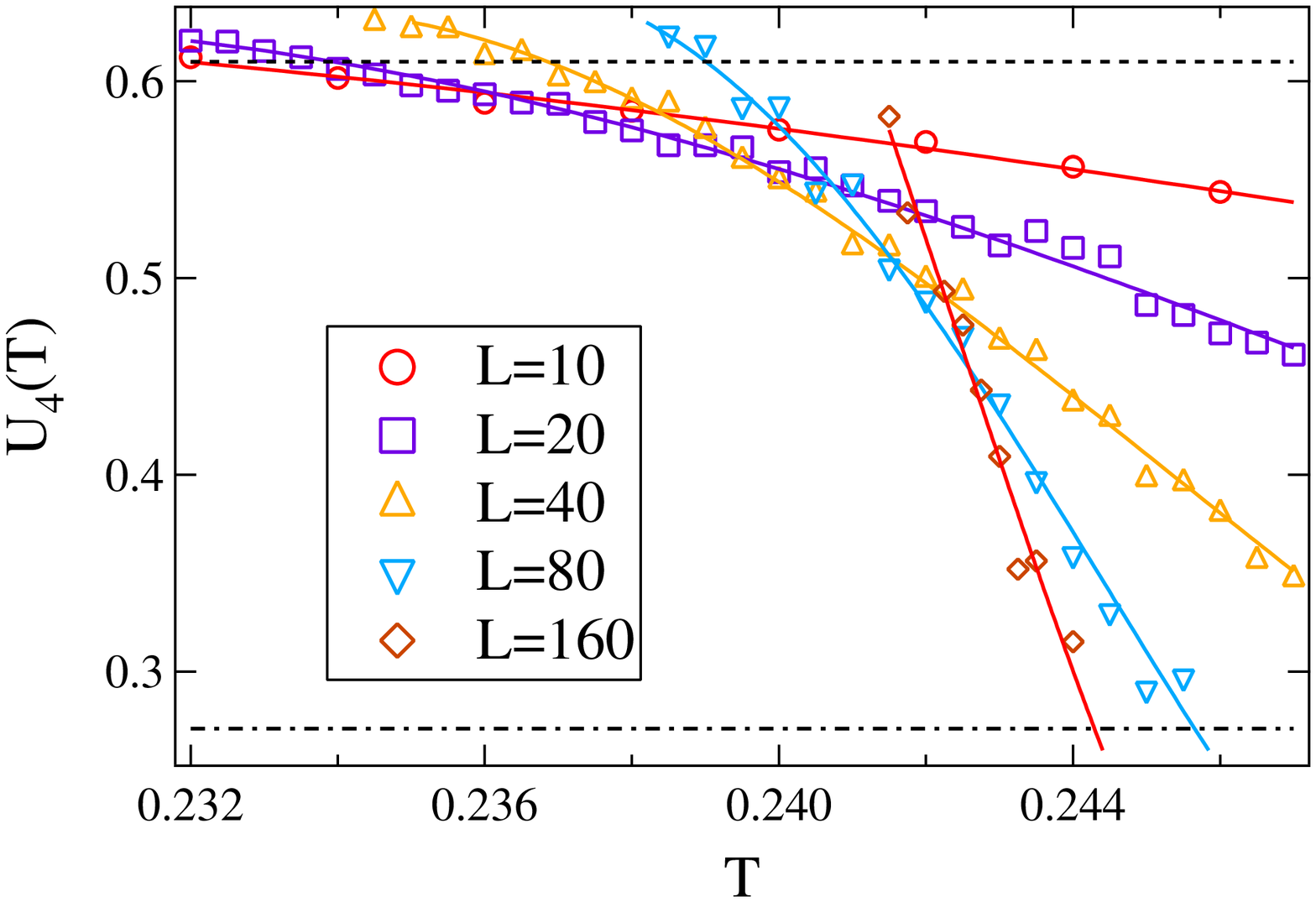} \\
({\rm a}) & ({\rm b})\\
\\
\includegraphics[scale=0.4]{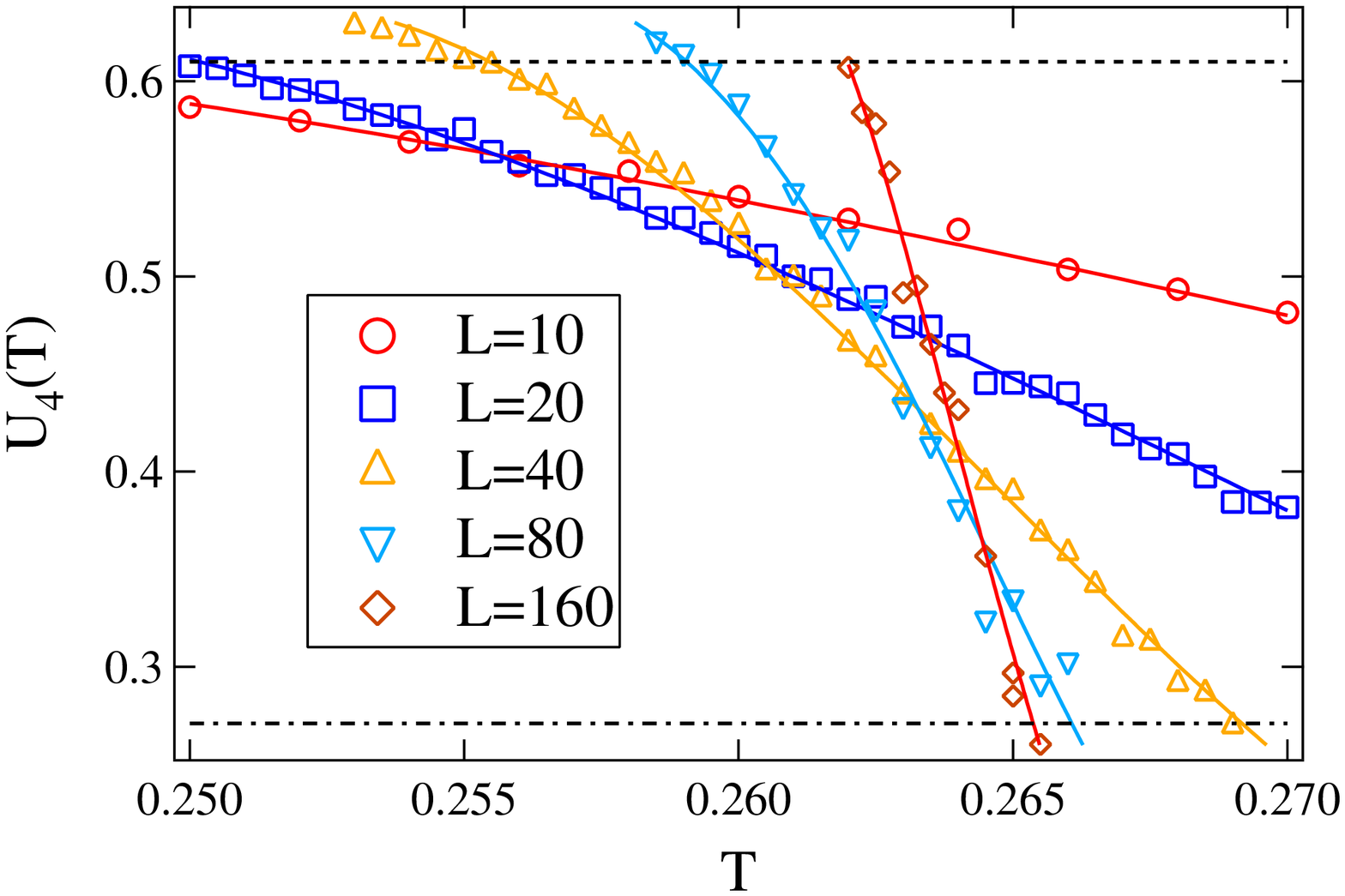} &
\includegraphics[scale=0.4]{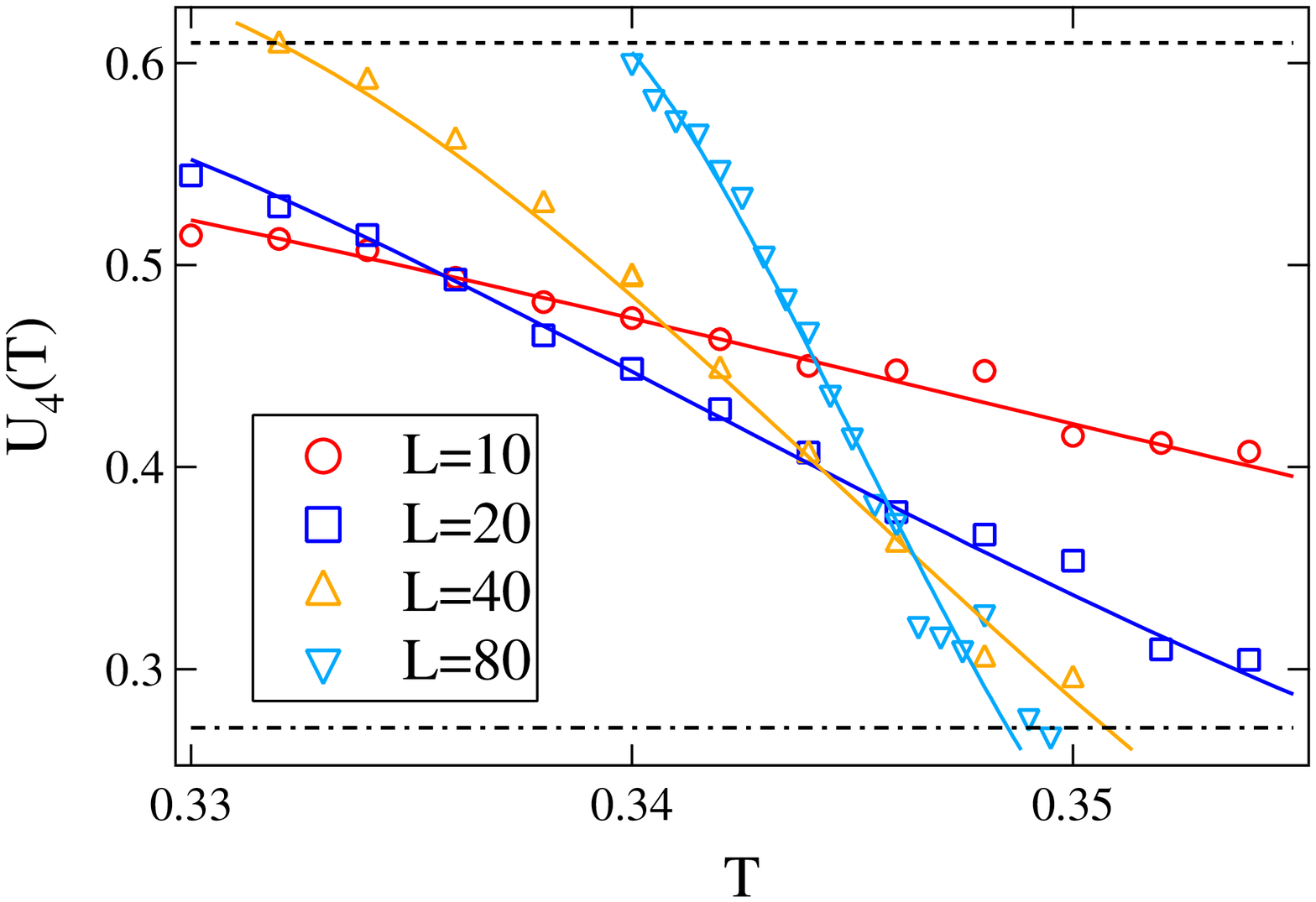} \\
({\rm c}) & ({\rm d})
\end{array}
$$
\caption{
(Color online) Temperature dependence of the Binder cumulant $U_4(L)$ for
(a) $R_{\rm H}=1.005$, (b) $R_{\rm H}=1.01$, (c) $R_{\rm H}=1.02$, and (d) 
$R_{\rm H}=1.05$.
Points denote Monte Carlo data, and the solid lines are polynomial fits. 
The upper and lower horizontal lines denote 
the fixed-point values for the Ising 
model $(U_4^{\rm IS}\simeq 0.61)$ and the infinite-range model $(U_4^{\rm MF}\simeq 0.27)$, respectively. 
The left vertical line in (a) represents 
the critical temperature of the pure Ising model. 
The linear system size $L$ is $10, 20, 40, 80, 160,$ and $320$ 
for circles, squares, upward triangles, downward triangles, diamonds, 
and asterisks, respectively.
}
\label{bindergraph}
\end{figure}

\begin{figure}[t]
$$\begin{array}{cc}
\includegraphics[scale=0.3]{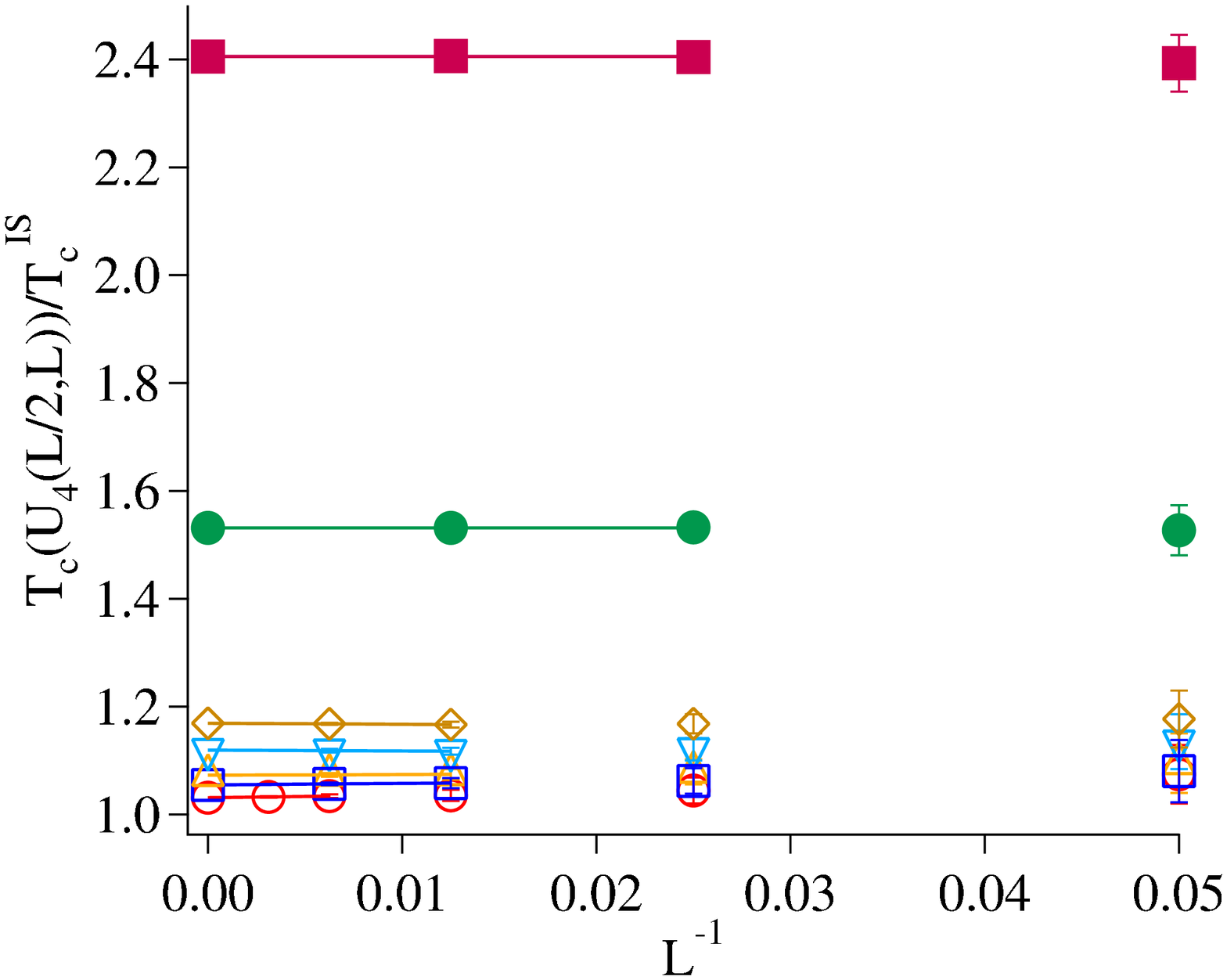} &
\includegraphics[scale=0.3]{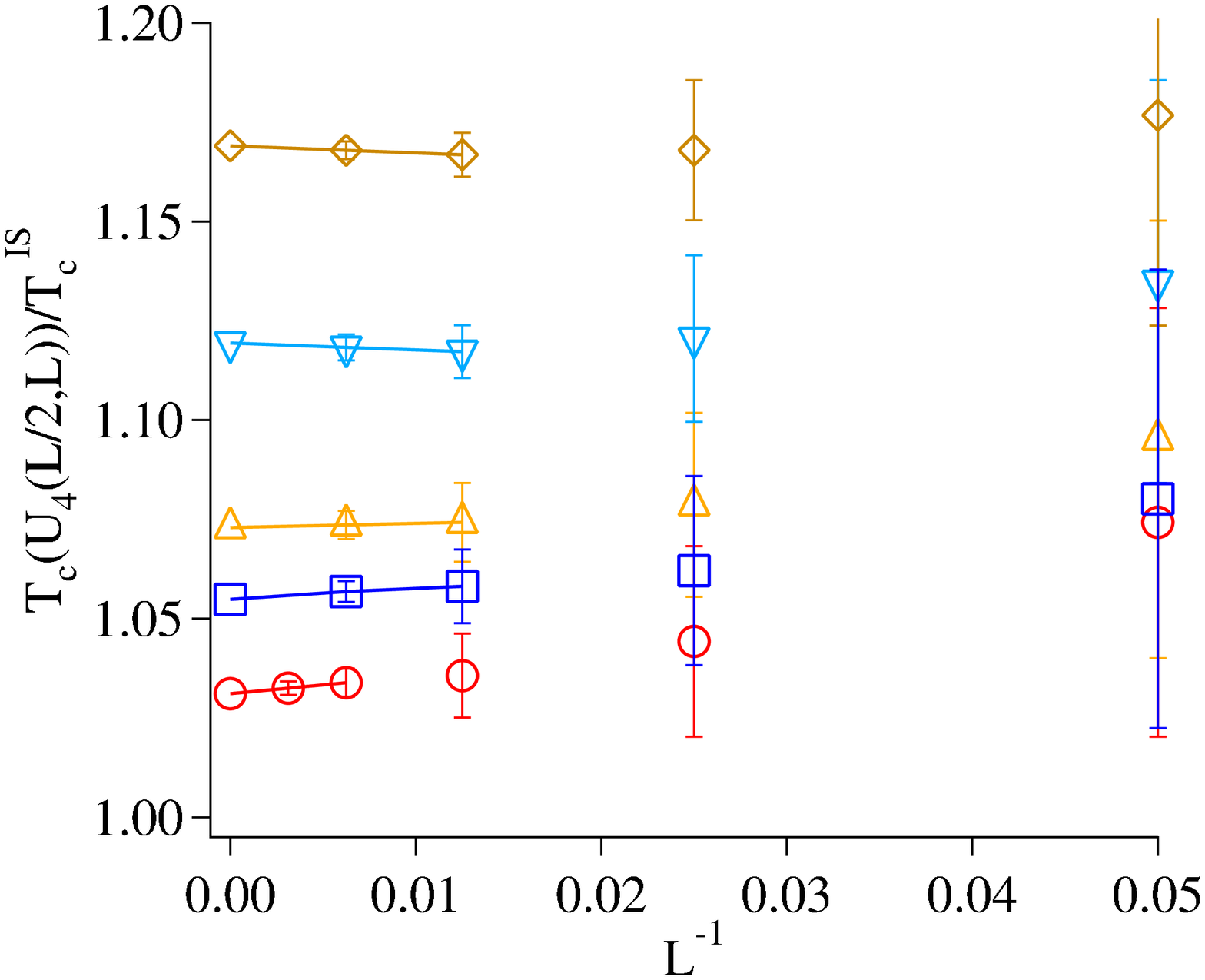} \\
({\rm a}) & ({\rm b}) \\
\end{array}
$$
\caption{(Color online) Size dependence of the crossing point of the Binder cumulant (a) 
and an enlarged view 
of the same for small $R_{\rm H}$ (b). 
From above to below, we show simulation results with $L=20,40,$ and $80$ for $R_{\rm H}=1.10$ and $1.05$, 
$L=20,40,80,$ and $160$ for $R_{\rm H}=1.02, 1.015, 1.01$ and $1.008$, 
$L=20,40,80,160$ and $320$ for $R_{\rm H}=1.005$. 
The points at $1/L=0$ are linear extrapolations from the two smallest nonzero values of $1/L$. 
Filled squares, filled circles, diamonds, downward triangles, upward triangles, open squares, and open circles 
represent 
$T_{\rm c}(L)$ for $R_{\rm H}=1.10$, 
$R_{\rm H}=1.05$, 
$R_{\rm H}=1.02$, 
$R_{\rm H}=1.015$, 
$R_{\rm H}=1.010$,
$R_{\rm H}=1.008$, 
and $R_{\rm L}=1.005$, respectively. 
}
\label{tcsize}
\end{figure}

In Fig.~\ref{bindergraph}, we see that the crossing points 
of the Binder cumulant decrease toward 
the mean-field fixed-point value, 
$U_{4}^{\rm MF}\simeq0.27\cdots$ \cite{Brezin,Luijten} 
from the Ising fixed-point value, $U_{4}^{\rm IS}\simeq 0.61 \cdots$ \cite{IsingCum} as $L$ increases. 
This indicates that a size dependent crossover occurs, and that the critical point of this model 
belongs to the mean-field universality class.  
We estimate the critical temperature as follows. 
Assuming that the critical behavior of the model belongs to the mean-field 
universality class, we get a series of upper bounds on the critical 
temperature as the temperature at which $U_4(L)$ 
crosses $U_4^{\rm MF}$. 
Lower bounds are given by the cumulant-crossing temperatures of $U_4(L)$ 
and $U_4(L/2)$. 
In Fig.~\ref{tcsize},  bars denote those upper bounds and lower bounds of the critical temperature, 
and we plot the middle points of those bounds by bullets. 
Increasing $L$, the temperature range 
between the upper bounds and the lower bounds becomes narrow. 
For each value of $R_{\rm L}/R_{\rm H}$, we extrapolate those middle points for the two 
largest system sizes versus $1/L$ to obtain the corresponding 
critical temperature in the thermodynamic limit. 
With this method, we expect that a critical temperature of a hybrid system with both short-range and long-range interactions 
can be obtained accurately in general. 
\begin{figure}[t]
$$
\includegraphics[scale=0.4]{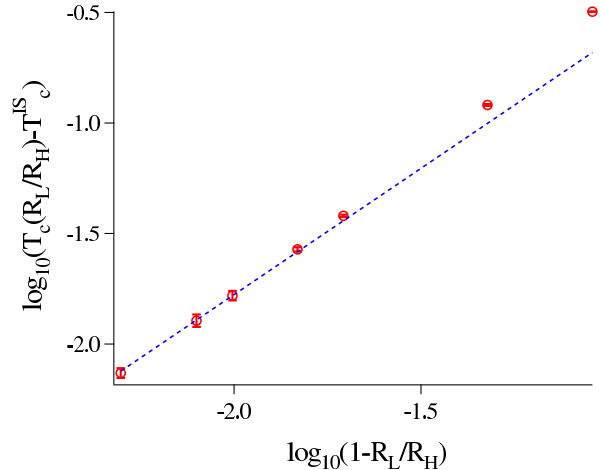}
$$
\caption{(Color online) The $\left(1-{R_{\rm L}\over R_{\rm H}}\right)$ dependence 
of the shift of the critical temperature 
in a log-log plot. 
The oblique dashed line (blue online) represents 
a numerical fit of the 4 leftmost points to $\left(1-{R_{\rm L}\over R_{\rm H}}\right)^{8\over 7}$.
}
\label{tcgraph}
\end{figure}
The data collapse well onto a straight line as shown in Fig.~\ref{tcgraph}, confirming 
the scaling form (\ref{tc}) for 
$R_{\rm H}$ smaller than $1.02$. 
The points fall above the line for 
large values of $1-R_{\rm L}/R_{\rm H}$. 
In this model, we fix the strength of the Ising interactions, 
while in the previous hybrid model~(\ref{model2}), the factor $1-\alpha$ multiplying the Ising Hamiltonian 
causes a deviation of the critical temperature in the opposite direction 
(see Fig.~4 of Ref.~1).

\section{Correlation length}
\label{secfour}
In this section, we perform MC simulations to 
test the scaling relation for the correlation length at the critical point~(\ref{xi}). 
We previously calculated the critical correlation length 
in the previous hybrid model (\ref{model2}) from MC simulations. \cite{Nakada} 
In that paper, we used 
the spin-correlation function $c(r)=\langle \sigma(r')\sigma(r'+r)\rangle$ 
to estimate the correlation length, excluding the contribution of long-range correlations from 
the correlation function. 
In the present model (\ref{model3}), 
because of strong anisotropy of the correlations, \cite{Miya1}
it is not practical to calculate the correlation length 
$\xi \left(R_{\rm L}/ R_{\rm H}\right)$ with this method.
\par
Here we instead obtain the correlation length from 
the structure factor, 
\beq
S({\mbold k})={1\over N}\sum_{l,m}\langle\sigma_l 
\sigma_m\rangle e^{i{\mbold k}\cdot{\mbold r}_{l,m}},
\eeq
which is readily measured in scattering experiments. 
All the contributions of the long-range correlations 
are given by $S({\mbold k}=0)$, 
so we can easily exclude them from the calculation. 
Here we note that there exist many experimental studies on the structure factors of magnetic materials, \cite{Ikeda}
obtained by neutron scattering. 
In the case of spin-crossover materials, the structures have been 
studied by single-crystal x-ray diffraction experiments. \cite{Guionneau, Lakhloufi} 
The structure factor for the HS/LS state domains discussed in the present paper 
can be obtained from diffuse x-ray scattering. 
\par
In the pure short-range Ising model at the critical temperature, 
the structure factor has its peak at ${\mbold k}=0$
 with infinitesimally narrow width in the thermodynamic limit.
However, in the hybrid model with both short-range and 
long-range interactions, 
the system prefers a spatially uniform configuration and 
the cluster size is suppressed even at the critical temperature, 
and the peak has a finite width in the thermodynamic limit. 
Here we obtain the correlation length at the critical point, $\xi_{\rm c}$ by calculating 
the characteristic peak width of the structure factor. 

\begin{center}
\begin{figure}[H]
$$\begin{array}{cc}
\includegraphics[scale=0.5]{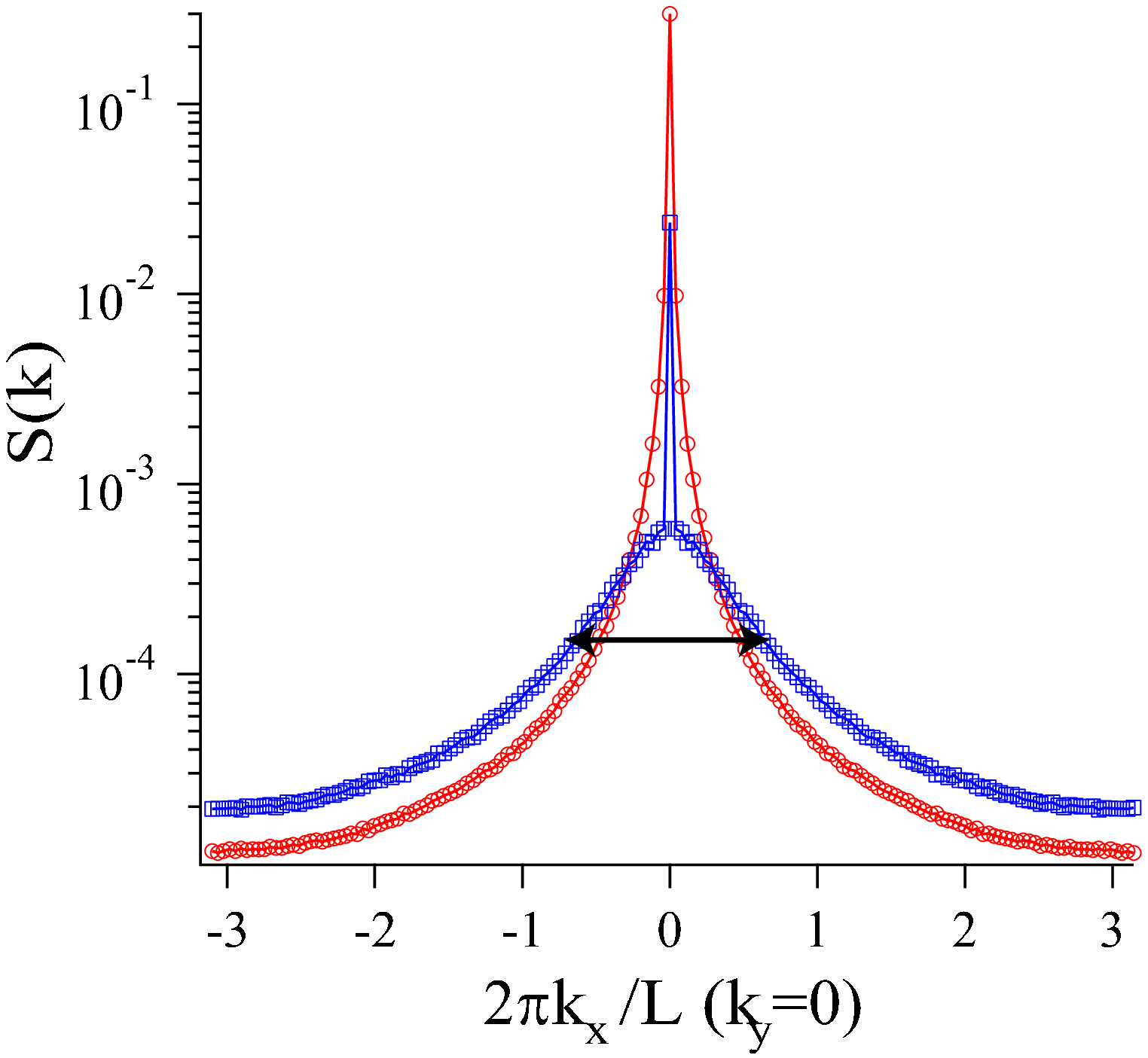} & 
\includegraphics[scale=0.5]{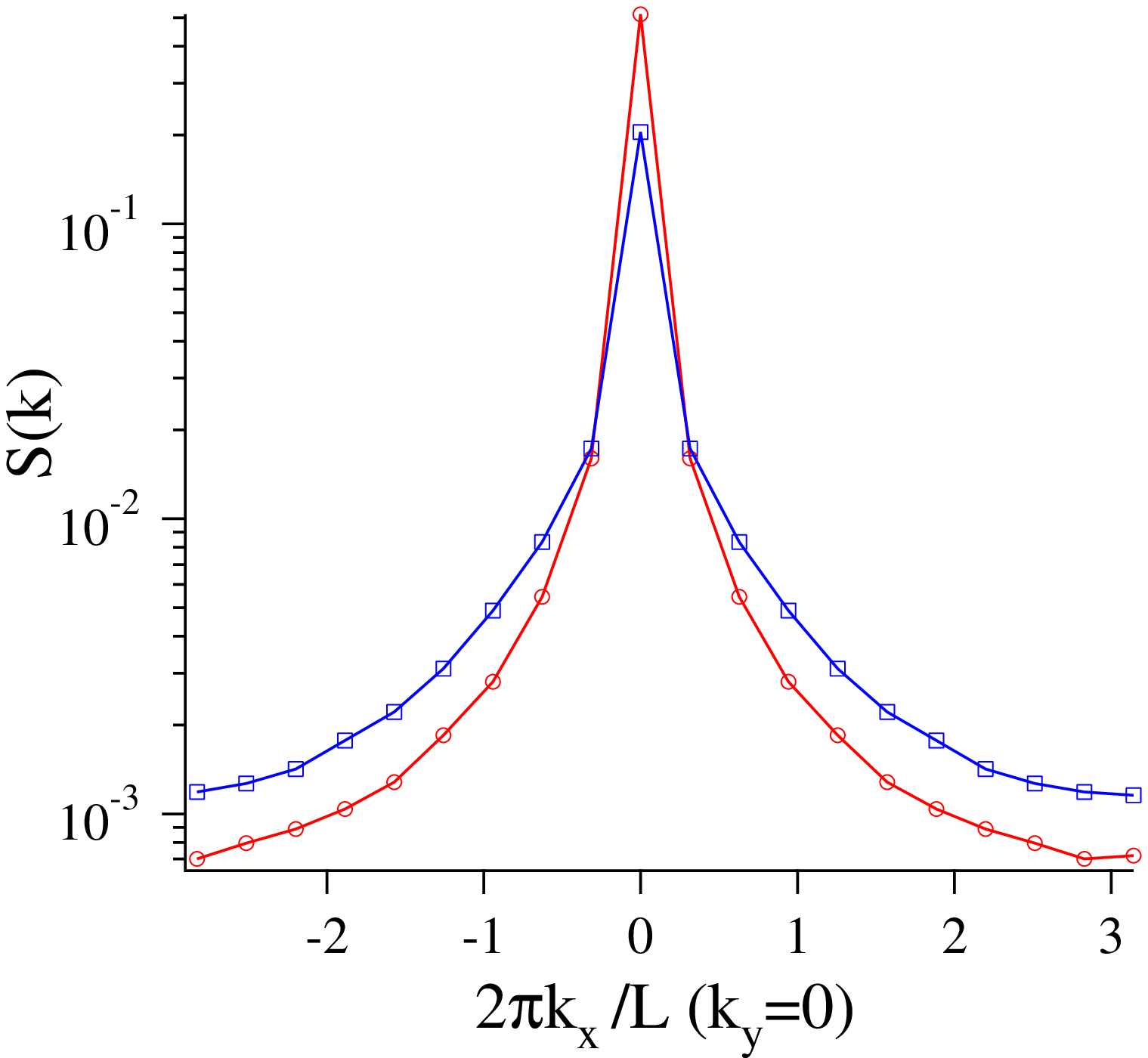} \\
({\rm a}) & ({\rm b}) \\
\end{array}$$
\caption{(Color online)
The structure factor of the pure short-range Ising model (circles) and 
the hybrid model with $R_{\rm H}=1.02$ (squares), at each its critical point. 
(a) $L=160$ and (b) $L=20$. }
\label{struc_snap}
\end{figure}
\end{center}

In Fig.~\ref{struc_snap}, we depict the structure factor of the pure Ising model at its critical temperature (circles)
together with that of 
the hybrid model at its critical temperature (squares). 
Figure~\ref{struc_snap} (a) shows the structure factor of the system for $L=160$, and Fig.~\ref{struc_snap} (b) is for $L=20$. 
We find a qualitative difference between these figures. 
Namely, we find a flat region in $S({\mbold k})$ of the hybrid model around ${\mbold k}={\mbold 0}$ in Fig.~\ref{struc_snap} (a), 
while $S({\mbold k})$ for both the pure short-range Ising model and the hybrid model show similar peaks in Fig.~\ref{struc_snap} (b). 
\par
The structure factor of the hybrid model consists of two parts:
A $\delta$-function at ${\mbold k}={\mbold 0}$ and a broad peak reflecting short-range order with the correlation length $\xi_{\rm c}$. 
The two-parts are superimposed in Fig.~\ref{struc_snap} (a). 
The flat region belongs to the diffuse peak, which is due to the finite cluster structure of the hybrid model. 
We find that for sufficiently large systems, 
the diffuse peak and the $\delta$-peak at ${\mbold k}={\mbold 0}$ are well distinguished. 
In those systems, we define the characteristic peak width $|{\mbold k}_{\rm peak}|$ 
as the spectral peak width of the diffuse peak as shown by the arrows in Fig.~\ref{struc_snap} (a). 
\par
In Fig.~\ref{struc_snap} (b), for small $L$, the resolution in ${\mbold k}$-space ($2\pi/L$) is rather coarse, 
so the data points do not reflect the finite width of the peak in the hybrid model well. 
It is hard to distinguish between the structure factors of the Ising model and the hybrid model in such small systems.
\par 
We note that the sum rule of the structure factor is $\sum_{\mbold k}S({\mbold k})=N$. 
In the hybrid model, $S({\mbold k})$ for large $|{\mbold k}|$ is larger than 
that of the pure short-range Ising model because $S({\mbold k})$ around ${\mbold k}={\mbold 0}$ is suppressed in the hybrid model. 
\par
We can estimate the correlation 
length $\xi$ from the structure factor by 
considering the first moment of $k^{-1}$,
\beq
\langle k^{-1}\rangle \equiv\sum_{k\neq 0}
{1\over | {\mbold k}|}S({\mbold k}),
\label{xical}
\eeq
where $k=| {\mbold k}|$. 
In our present model, the correlation function of the spin configuration is given by \cite{Miya1, Konishi_D}  
\beq
\langle \sigma_i \sigma_j \rangle = c({\mbold r}_{i,j})=
c^{\rm SR}(|{\mbold r}_i - {\mbold r}_j |)
+c^{\rm LR}.
\eeq
Here, $c^{\rm SR}(|{\mbold r}_i - {\mbold r}_j |)$ denotes the contributions from the short-range interactions, and 
$c^{\rm LR}$ denotes those from the long-range interactions. 
We note that the long-range correlation $c^{\rm LR}$ does not depend on the distance $|{\mbold r}_i - {\mbold r}_j |$. 
In Eq.~(\ref{xical}), the contribution from the long-range correlations $c^{\rm LR}$ (i.e., $S({\mbold k}={\mbold 0})$) is excluded, 
so we use the asymptotic formula 
for the pure short-range Ising model, \cite{T.T.Wu}
\beq
c(r)\propto {1\over r^{2-d+\eta}}e^{-r/\xi}.
\label{correIsing}
\eeq
In the thermodynamic limit, we replace the sum in (\ref{xical}) 
by an integral: 
\begin{eqnarray}
\langle k^{-1} \rangle&=&\int_{k=2\pi/L}^{\infty}\int_{\phi=0}^{2\pi}kdkd\phi{1\over k}\int_{r=0}^{\infty}
\int_{\theta=0}^{\rm 2\pi}rdrd\theta e^{ikr{\rm cos}\theta}c(r)\\
&\propto&\int_{k=0}^{\infty}dk\int_{r=0}^{\infty}dr
r^{1-\eta}e^{-r/\xi}J_0(kr)\\ 
&=&\xi^{1-\eta}\int_{K=0}^{\infty}\int_{R=0}^{\infty} 
R^{-\eta}e^{-R}J_0(K)dKdR\\
&\propto&\xi^{1-\eta}=\xi^{3/4}.
\label{correIsingint}
\end{eqnarray}
In the last line, we use the substitutions, $K=kr$, $R=r/\xi$, 
and $J_0(K)$ is the Bessel function. 
There is no singularity in the integral, and $\langle k^{-1}\rangle$ gives a power 
of the correlation length. 
Therefore we define the critical correlation length $\xi_{\rm c}$ at the critical temperature $T_{\rm c}$ as follows, 
\beq
\xi_{\rm c}\propto \langle k^{-1}\rangle^{4\over 3}_{\rm c},
\label{def_xic} 
\eeq
where $\langle k^{-1}\rangle_{\rm c}$ represents the value at the critical point. 
The scaling relation for $\langle k^{-1}\rangle_{\rm c}$ is, by making use of 
(\ref{xi}), 
\beq
\langle k^{-1}\rangle_{\rm c}^{4\over 3} \propto \xi_{\rm c}
=
Lf\left(L\left(1-{R_{\rm L}\over R_{\rm H}}\right)^{8\over 7}\right),
\label{scalexi}
\eeq
where $g(x)\propto {1\over x}$ for $x \rightarrow \infty$.

\begin{figure}[H]
$$\begin{array}{cc}
\includegraphics[scale=0.4]{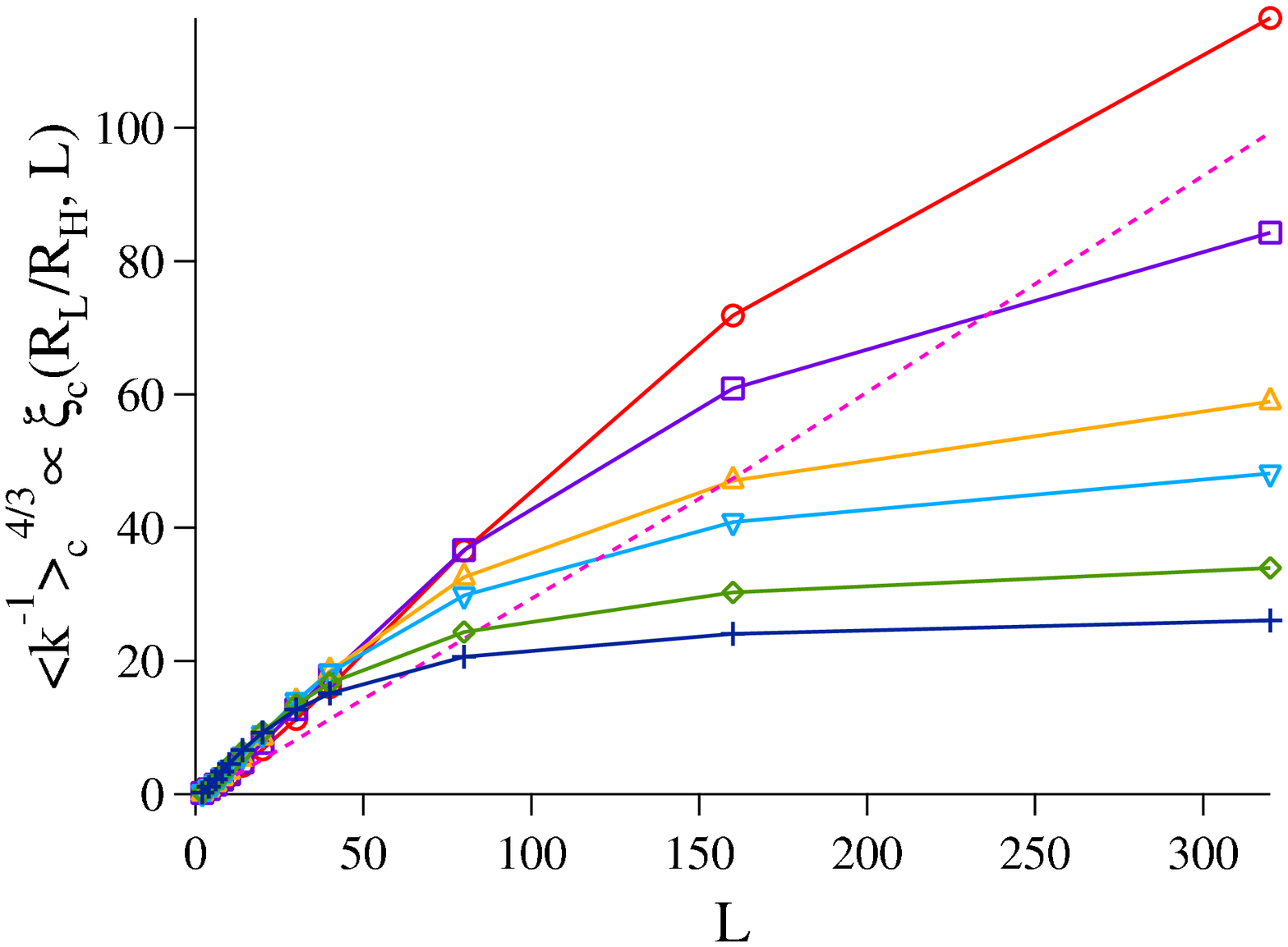} & 
\includegraphics[scale=0.4]{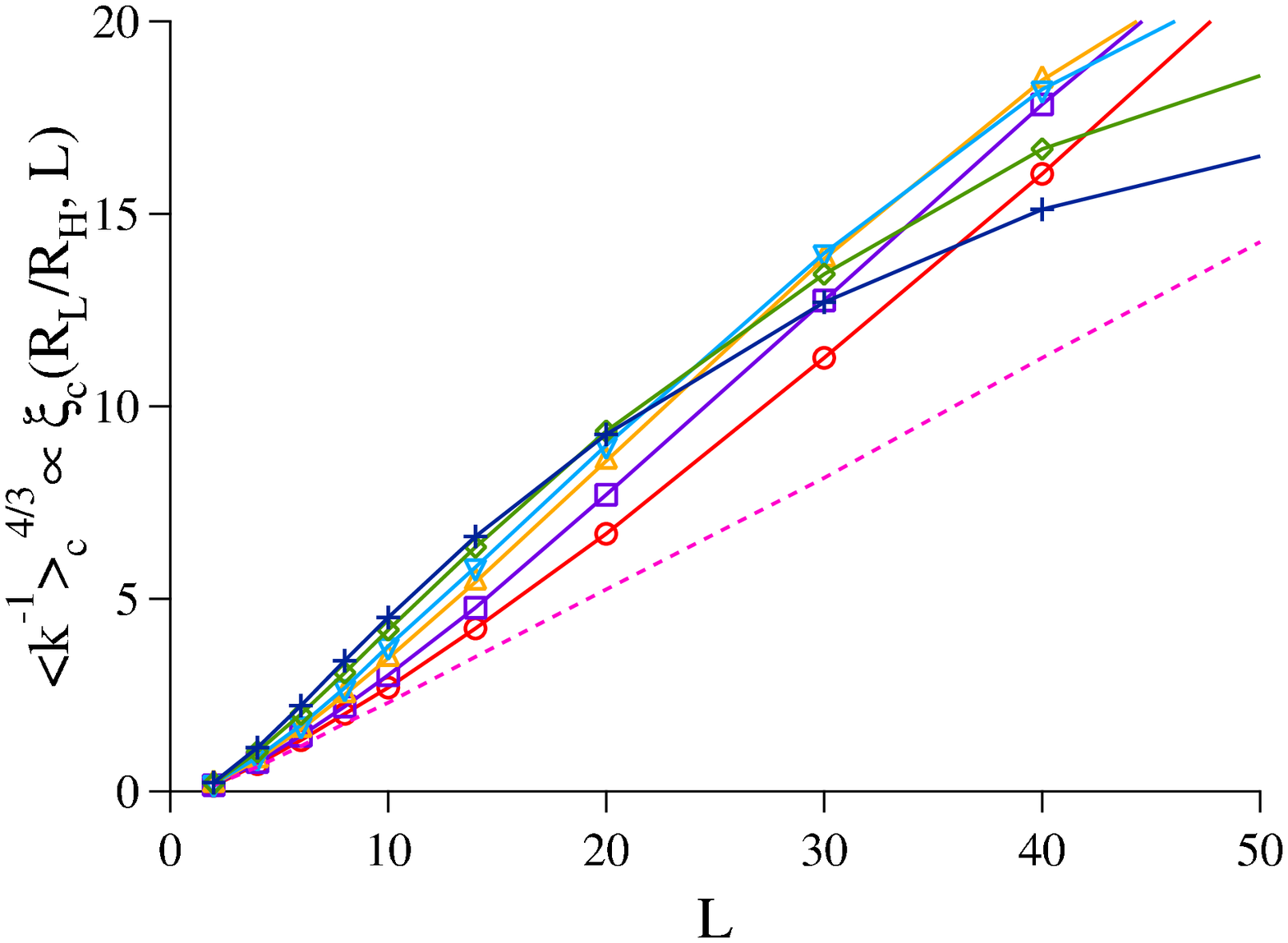} \\
({\rm a}) & ({\rm b}) \\
\end{array}$$
\caption{(Color online) The size dependence of 
$\langle k^{-1}\rangle^{4\over 3}_{\rm c}\propto \xi_{\rm c}$ 
at the critical point $T_{\rm c}(R_{\rm L}/R_{\rm H})$ (a), 
and an enlarged view of the same for small $L$ (b). 
$L=2,4,6,8,10,14,20,30,40,80,160$ and $320$. 
Circles, squares, upward triangles, downward triangles, diamonds 
and crosses 
represent 
$T=0.23107$ for $R_{\rm H}=1.003$, 
$T=0.23428$ for $R_{\rm H}=1.005$, 
$T=0.23965$ for $R_{\rm H}=1.008$, 
$T=0.24345$ for $R_{\rm H}=1.01$,
$T=0.2537$ for $R_{\rm H}=1.015$ 
and $T=0.2649$ for $R_{\rm L}=1.02$, respectively. 
The dashed line (blue online) represents 
$T=0.2269$ for $R_{\rm H}=1.00$, the pure Ising-model limit. 
}
\label{xiL}
\end{figure}

We obtained $\langle k^{-1}\rangle_{\rm c}$ 
and the corresponding values of $\xi_{\rm c}$ 
for 
various values of $L$ and $R_{\rm L}/R_{\rm H}$. 
Those are plotted in 
Fig.~\ref{xiL}. 
In the enlarged view of Fig.~\ref{xiL} for small $L$, we find inflection points. 
For small $L$, $\xi_{\rm c}(R_{\rm L}/R_{\rm H}, L)$ grows faster in the hybrid model than in the pure Ising model. 
\par
Here let us consider the reason for this behavior. 
This fast growth is due to the coarse resolution in ${\mbold k}$-space. 
The characteristic peak width of the hybrid model does not represent well the structure factor in the small systems (Fig.~\ref{struc_snap}(b)), and the correlation length of the hybrid model is overestimated in Eq.~(\ref{scalexi}). 
This extra size dependence for small size comes from the discontinuous nature of ${\mbold k}$-space, and is not essential for the present purpose. 
\par
For large $L$, the growth of $\xi_{\rm c}(R_{\rm L}/R_{\rm H}, L)$ becomes much slower than in the pure Ising model. 
For $R_{\rm H}=1.02 $ and $L=320$, $\xi_{\rm c}(R_{\rm L}/R_{\rm H}, L)$ is almost saturated. 
For other values of $R_{\rm H}$, $\xi_{\rm c}(R_{\rm L}/R_{\rm H}, L)$ is expected to saturate for still larger systems. 
It means that for sufficiently large systems, the difference of the structure factor of the hybrid model and the pure short-range model is clear. 

The same behavior is found in the previous hybrid model (\ref{model2}). 
In both models, the previous hybrid model (\ref{model2}) and the present model (\ref{model3}), 
the long-range interactions suppress the clustering of spins. 
Thus there appear tightly correlated effective block spins. 
In the coarse-grained Hamiltonian with the block spins, the long-range interactions become 
effectively stronger than the short-range interactions. \cite{Nakada}

According to this picture, for $L$ smaller than the clusters caused by the short-range interaction, the long-range interaction is irrelevant. 
Thus, for small $L$, the system is an effective short-range interacting system. 
On the other hand, for sufficiently large $L$, the system is an effective long-range interacting system. 
We refer the crossover length as $L_{\rm cl}$, which we define as 
$L_{\rm cl}=1/|{\mbold k}_{\rm peak}|$.
The inflection points in Fig.~\ref{xiL} also indicate the crossover length of the system size, $L_{\rm cl}$.
For $L > L_{\rm cl}$, 
the correlation length at the critical point, $\xi_{\rm c}(R_{\rm L}/R_{\rm H}, L)$, tends to be saturated. 
For $L < L_{\rm cl}$, $\xi_{\rm c}(R_{\rm L}/R_{\rm H}, L)$ grows faster than linear. 
\par
We also found that the peak position of ${\tilde \chi}(L)\equiv {1\over N}(\langle M^2 \rangle -\langle |M| \rangle ^2)$ shows similar behavior. 
The peak position, the effective `critical point' for a system of size $L$ , saturates at the critical temperature in the thermodynamic limit:
\beq
T_c\left({R_{\rm L}\over R_{\rm H}},L\right)\rightarrow T_{\rm c}\left({R_{\rm L}\over R_{\rm H}},\infty\right).
\eeq
As in the previous work, \cite{Nakada} we found a non-monotonic dependence of the peak position as a function of $L$. 
For small $L$, the peak positions approach $T_{\rm c}$ from the high-temperature-side, while for large $L$, the peak position is on the low-temperature-side and eventually 
approaches $T_{\rm c}$ from below. This non-monotonic behavior indicates the crossover from the effective short-range system to the effective long-range system. 
It is an interesting problem to study the relation between the crossover phenomena in ${\tilde \chi}(L)$ and the correlation length $\xi_{\rm c}$ in Fig.~\ref{xiL}.  
We expect that this general system size dependence occurs in any hybrid model 
where effective spins play an important role. 
\par
\begin{figure}[H]
$$
\includegraphics[scale=0.5]{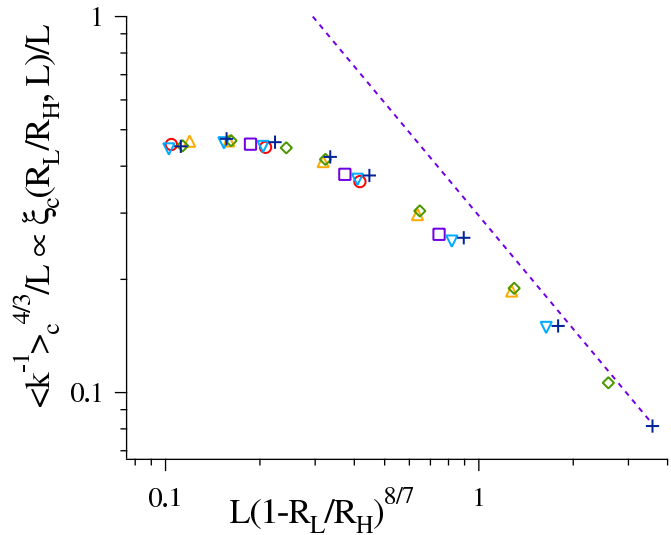}
$$
\caption{(Color online) Scaling plot of 
the correlation length at the critical point. 
Circles, squares, upward triangles, downward triangles, diamonds, 
and crosses represent 
$T=0.23107$ for $R_{\rm H}=1.003$,
$T=0.23428$ for $R_{\rm H}=1.005$, 
$T=0.23965$ for $R_{\rm H}=1.008$, 
$T=0.24345$ for $R_{\rm H}=1.01$,
$T=0.2537$ for $R_{\rm H}=1.015$, 
and $T=0.2649$ for $R_{\rm L}=1.02$, respectively. 
Data are included only for $L>L_{\rm cl}$. 
The dashed line is proportional to $y=x^{-1}$. 
The data are in good agreement with the scaling relation, 
but they converge more gradually to the asymptotic formula than when the correlation length is calculated from the 
correlation function. \cite{Nakada} 
}
\label{xiscale}
\end{figure}

In appendix A, we discuss in detail how the correlation lengths are measured. 
In Fig~\ref{xiscale}, we plot the data for $L>L_{\rm cl}(R_{\rm L}/R_{\rm H})$ in a 
scaling plot of the form (\ref{scalexi}). 
We find that the data collapse onto a scaling function and 
thus we conclude that (\ref{scalexi}) is justified. 
\section{Summary and further discussion}
\label{secfif}

In general, spin-crossover and related materials have both 
short-range and long-range interactions. 
If the elastic potential is chosen spin-state dependent 
\cite{Nicolazzi, Nicolazzi2}, in which different 
potential functions (coupling constants) are given for 
LS-LS, LS-HS, and HS-HS molecular pairs, 
it is considered that the interaction has intrinsically both 
short-range and long-range components.
In this paper, we have studied critical properties of the present model (\ref{model3}) 
with elastic and short-range interactions. 
By Monte Carlo simulations we confirmed that the present model obeys a scaling 
relation for the shift of the critical temperature as a function of the strength of the long-range 
interaction, which is determined by the ratio of the radii of the molecules in both states. 
We similarly confirmed that the present model also obeys a scaling relation for 
the correlation length at the critical point, which was found in our 
previous work for the hybrid model with infinite-range and nearest-neighbor interactions. \cite{Nakada} 
Although the origin of the long-range interaction is very different in the previous model and the present model, 
we found several kinds of universality in their critical properties. 
\par
In the present model, because of the anisotropy in the correlation function, 
we estimated the correlation length from the structure factor. 
By this method, the correlation length can be measured in scattering experiments on real materials. 
Because spin-crossover materials usually undergo first-order phase transitions, experiments on the structure factors of these materials have been so far only done for HS/LS ordered phase. \cite{Guionneau, Lakhloufi} 
Our results show that the structure factor of spin-crossover materials should exhibit a peak with finite width at the critical point due to the elastic interaction. 
We hope such unique behavior will be observed by single crystal x-ray diffraction along the coexistence line and  
at the critical point. 
\par
It was confirmed that the present model (\ref{model3}) 
possesses an effective long-range 
interaction. 
The details of the effective long-range interactions introduced by 
the elastic degrees of freedom in the present model are not known. 
Only for $d=1$, \cite{Boukheddaden}
it has been shown rigorously that the model can be mapped 
onto an short-range Ising model. 
There has also been much previous research on three-dimensional 
elastic solids, and it is generally argued that the 
dominant long-range interactions are of a dipole-dipole nature, 
$\sim 1/r^3$. \cite{Stewart}
Although the effects of distortions in the present model are not identical 
to those in the classical elastic media for which these results were obtained, 
we assume that the elastically mediated interactions in our model also 
are of such long-range type. 
However, in our present model the spin correlation
 function shows infinite-range correlations above the 
critical temperature. \cite{Miya1, Konishi_D}
Therefore, we assume that there are similar but unknown infinite-range interactions 
in the present model. 
Understanding the mechanism by which the infinite-range interactions arise remains 
an intriguing problem for future research.

\section*{Acknowledgments}
The present work was partially supported by the Mitsubishi Foundation, 
Grant-in-Aid for Scientific Research on Priority Areas, KAKENHI (C)  23540381, 
and also the Next Generation 
Super Computer Project, Nanoscience Program from MEXT of Japan. 
T.M. acknowledges the support from JSPS (Grant No. 227835). 
The numerical calculations were supported by the supercomputer center of
ISSP of University of Tokyo. Work at Florida State University was supported in 
part by U.S. National Science Foundation Grants No. DMR-0802288 and DMR-1104829.

\appendix
\section{Calculating the correlation length $\xi$}
\label{app}
In the previous work on the model defined in~(\ref{model2}), there exist both 
short-range and infinite-range interactions, but no anisotropy. 
Therefore we could exclude the contribution of 
the long-range correlations uniformly from the correlation function, 
obtaining the correlation length from the following relation,
\beq
\xi={\int_{0}^{L/2}\left( c(r)-c(L/2) \right)rdr \over 
\int_{0}^{L/2}\left( c(r)-c(L/2) \right)dr}.
\label{xicalprevious}
\eeq
Here, $c(L/2)$ which is proportional to ${1\over \sqrt[]{N}}$ is a 
good approximation for the large-$r$ limit of $c(r)$. 
\par
In the present paper, we adopt another method because of the strong 
anisotropy in the correlation function. 
In a $d-$dimensional system, we can estimate $\xi$ from
\beq
\langle k^{1-d}\rangle\equiv\sum_{k\neq 0}
{1\over | {\mbold k}| ^{d-1}}S({\mbold k}) 
\propto \xi^{2d-3-\eta}.
\label{another}
\eeq
In the present model defined by~(\ref{model3}) with 
very weak elastic interactions, 
there exist rather large, but finite clusters 
at the critical point due to the short-range interactions. 
In (\ref{another}), we sum the structure factor with 
importance $1/| {\mbold k}| ^{d-1}$ in order to avoid divergences and to collect the 
contributions from length scales on the order of $L$, using the asymptotic form of the 
correlation function~(\ref{correIsing}).

\end{document}